\documentclass[journal,twoside]{IEEEtran}
\usepackage{amssymb}
\usepackage{amsmath}
\usepackage{amsfonts}
\usepackage{empheq}
\usepackage{color}

\newcommand{\beq}{\begin{equation}}
\newcommand{\eeq}{\end{equation}}
\newcommand{\bea}{\begin{eqnarray}}
\newcommand{\eea}{\end{eqnarray}}

\newcommand{\comment}[1]{}
\renewcommand{\d}{{\rm d}}

\ifCLASSINFOpdf
  \usepackage[pdftex]{graphicx}
\else
  \usepackage[dvips]{graphicx}
\fi

\begin{document}
%
\title{Conditions for Parametric and Free-Carrier Oscillation in Silicon Ring Cavities}
%
%
%

\author{Ryan~Hamerly, 
        Dodd~Gray, 
        Christopher~Rogers, 
        and~Kambiz~Jamshidi~\IEEEmembership{Member,~IEEE} 
\thanks{R.~Hamerly was with the National Institute of Informatics, Tokyo, 101-8403 Japan.  He is now with the Research Laboratory of Electronics at the Massachusetts Institute of Technology, Cambridge, MA, 02139 USA (email: rhamerly@mit.edu)}%
\thanks{D.~Gray and C.~Rogers are with the E.~L.~Ginzton Laboratory, Stanford University, Stanford, CA, 94305 USA (email:dodd@stanford.edu, cmrogers@stanford.edu)}%
\thanks{K.~Jamshidi is with the Integrated Photonic Devices Group, Technische Universit\"{a}t Dresden, Dresden, 01062 Germany (email: kambiz.jamshidi@tu-dresden.de)}%
\thanks{Manuscript received January 31, 2018, revised March 30, 2018}}

%
%

\markboth{Journal of Lightwave Technology,~Vol.~xx, No.~x, xxxx~2018}%
{Hamerly \MakeLowercase{\textit{et al.}}: Conditions for Parametric and Free-Carrier Oscillation in Silicon Ring Cavities}
%



\maketitle

\begin{abstract}
We model optical parametric oscillation in ring cavities with two-photon absorption, focusing on silicon at 1.55$\mu$m.  Oscillation is possible if free-carrier absorption can be mitigated; this can be achieved using carrier sweep-out in a reverse-biased p-i-n junction to reduce the carrier lifetime.  By varying the pump power, detuning, and reverse-bias voltage, it is possible to generate frequency combs in cavities with both normal and anomalous dispersion at a wide range of wavelengths including 1.55$\mu$m.  Furthermore, a free-carrier self-pulsing instability leads to rich dynamics when the carrier lifetime is sufficiently long.
\end{abstract}

\begin{IEEEkeywords}
Silicon photonics; Frequency comb; Nonlinear optics.
\end{IEEEkeywords}

%
\IEEEpeerreviewmaketitle

\section{Introduction}

\IEEEPARstart{O}{ptical} parametric amplification (OPA) and oscillation (OPO) are useful phenomena in both bulk and integrated optics, and are employed in numerous applications including frequency-comb generation \cite{Pasquazi2018}, optical logic \cite{Inagaki2016}, and quantum information \cite{Roslund2014}.  Frequency combs generated from parametric oscillators can also be used for wavelength-division multiplexing \cite{Pfeifle2014} and as a source for tunable microwave or terahertz radiation \cite{Kitayama1997}.  The development of small mode-volume resonators with ultra-high $Q$ factors has opened a new path to utilizing parametric processes \cite{Kippenberg2004}, and has led to successful frequency-comb demonstrations in many platforms including MgF$_2$/CaF$_2$ cavities \cite{Herr2014}, SiO$_2$ microtoroids \cite{Chembo2010}, and Si$_3$N$_4$ microrings \cite{Saha2013}.  However, the weak optical nonlinearities of these materials require extremely low-loss structures (cavity $Q \gtrsim 10^6$--$10^8$ depending on the platform) to realize the nonlinearity at reasonable power levels, which poses challenges for device fabrication and yield \cite{Pfeiffer2016} 

Silicon is a promising alternative because of its large nonlinearity ($20\times$ higher than Si$_3$N$_4$ and $200\times$ higher than SiO$_2$ \cite{Foster2006}), high refractive index permitting smaller mode volumes, and established CMOS-compatible fabrication process.  These advantages significantly relax the power and loss constraints for comb generation relative to other platforms.  However, at wavelengths shorter than 2.3$\mu$m, silicon devices suffer from two-photon absorption (TPA) and the resultant free-carrier absorption (FCA), which compete with the parametric process and suppress gain.  Nevertheless, OPA has been demonstrated in pulsed mode at 1.55$\mu$m in silicon waveguides \cite{Foster2006}, and mid-IR gain \cite{Liu2010} and comb generation \cite{Griffith2015} have been reported beyond the TPA cutoff, indicating that gain is possible as long as FCA can be controlled.

\begin{figure}[tbp]
\begin{center}
\includegraphics[width=0.92\columnwidth]{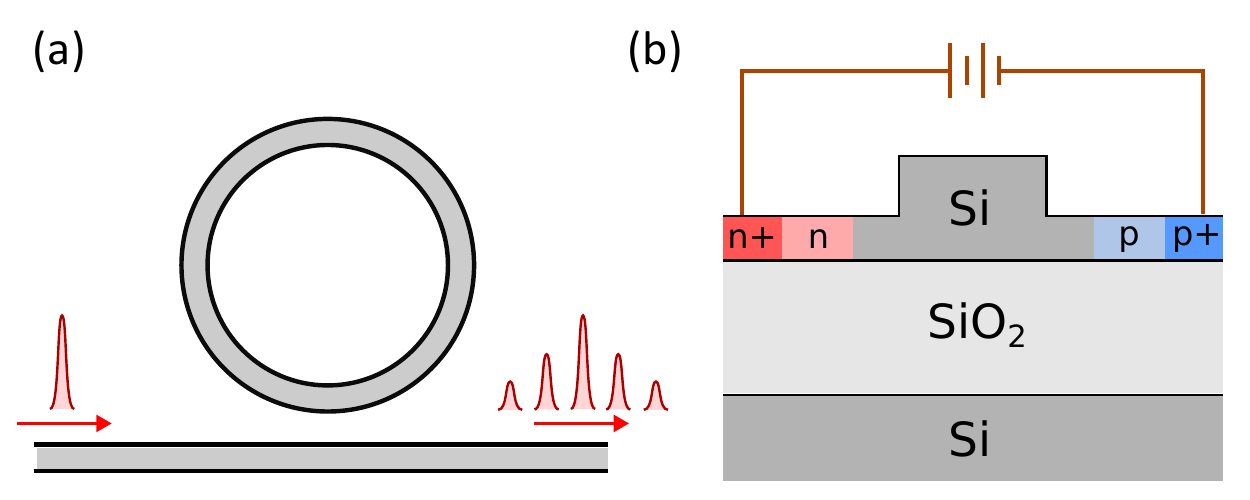}
\caption{(a) Illustration of the system: continuous-wave light enters the cavity and produces an output at multiple frequencies. (b) Cross-section of the device: an optical rib waveguide inside a reverse-biased p-i-n diode, used for free-carrier sweep-out.  From Ref.~\cite{HamerlyMWP}.}
\label{fig:mwp-f1}
\end{center}
\end{figure}

This paper studies the conditions for parametric oscillation and comb generation in silicon in the near-IR, and the constraints imposed by TPA and FCA.  TPA slightly degrades performance but by itself does not prevent oscillation, despite silicon's low nonlinear figure of merit \cite{Dinu2003}.  To overcome FCA, one needs to simultaneously minimize loss and free-carrier lifetime.  While ion implantation and reducing waveguide dimensions can reduce the carrier lifetime, this comes at the expense of higher loss \cite{Dekker2007}.  Active carrier removal, by contrast, exploits a strong reverse-bias field in a p-i-n junction (Fig.~\ref{fig:mwp-f1}) to reduce the carrier lifetime considerably while maintaining a low optical loss \cite{TurnerFoster2010}.  For a 220$\times$450 nm waveguide, the lifetime varies between 5--100 ps, tunable by the bias voltage, and is 1--2 orders of magnitude smaller than the $\sim$1 ns lifetimes observed in ordinary waveguides \cite{Gajda2011}.  



In Sec.~\ref{sec:mm}, we introduce the mathematical model consisting of a Lugiato-Lefever equation with free-carrier terms.  Sec.~\ref{sec:gain} treats the case of a continuous-wave pump and derives the conditions for parametric gain.  These conditions are studied in both anomalous- and normal-dispersion rings, focusing on silicon at 1.55$\mu$m.  Sec.~\ref{sec:sim} presents numerical simulations of comb formation in both the anomalous- and normal-dispersion regimes to study the effects of TPA / FCA on the comb.  Finally, Sec.~\ref{sec:lamdep} generalizes our discussion to the conditions for net gain at arbitrary wavelengths.  Gain is possible at 1.55$\mu$m with waveguide losses $\lesssim 2$ dB/cm and carrier lifetimes $\lesssim 100$ ps, and these bounds become are much more lenient at longer wavelengths.

\section{Mathematical Model}
\label{sec:mm}

To model this system, we use a normalized Lugiato-Lefever equation (LLE) \cite{Lugiato1987} with additional degrees of freedom for the free carriers \cite{Hamerly2017}: 
\bea
	\frac{\partial \bar{a}}{\partial\bar{\tau}} & \!=\! & \Bigl[(-1 - i\bar{\Delta}_0 ) + (i - r)|\bar{a}|^2 \nonumber \\
	& & \ \ \ \ +\, (-i - \mu^{-1}) \bar{n}_c - i\,\eta \frac{\partial^2}{\partial \bar{t}^2}\Bigr] \bar{a} 
	+ \bar{S} \label{eq:da-sm2} \\
	\frac{\d \bar{n}_c}{\d\bar{\tau}} & \!=\! & \frac{2}{\tau_c/\tau_{\rm ph}} \left[\bar{\chi}_c \left(\frac{1}{t_R}\int_0^{t_R}{|\bar{a}|^4\d t}\right) - \bar{n}_c\right] \label{eq:dn-sm2}
\eea
The dynamical variables are the (normalized) optical field $\bar{a}(\bar{t},\bar{\tau})$ and the carrier density $\bar{n}_c(\bar{\tau})$.  The slow time $\tau$ is normalized to the photon lifetime $\tau_{\rm ph}$: $\tau = 2\tau_{\rm ph}\bar{\tau}$; the fast time $t = \xi_t \bar{t}$ is scaled to set the dispersion term to $\pm 1$, and the field $a = \xi_a \bar{a}$ and carrier density $n_c = \xi_n \bar{n}_c$ are normalized to the strength of the Kerr and free-carrier dispersion, respectively
.  $\bar{\Delta}_0 = 2\tau_{\rm ph}(\omega_{\rm cav} - \omega_{\rm p})$ is the cold-cavity detuning, the constants $r=0.2$ and $\mu=25$ are material parameters \cite{Lin2007}, $\eta\in\{-1,+1\}$ is the sign of the waveguide group-velocity dispersion (GVD), and $\bar{S}$ is the (normalized) input field.  The carrier lifetime $\tau_c$ and free-carrier dispersion term $\bar{\chi}_c$ (given by $\bar{\chi}_c = (\tau_c/\tau_{\rm ph})(r\mu\sigma/4\hbar\omega\gamma v_g) \approx 5\,\tau_c/\tau_{\rm ph}$) can be tuned by the reverse bias voltage.  The cavity round-trip time and group velocity are $t_R$ and $v_g$, while $\gamma = 2\pi n_2/\lambda$ and $\sigma = \d\alpha/\d n_c$ are the Kerr and FCA coefficients.  See Appendix \ref{sec:app-lle} for details.  Eqs.~(\ref{eq:da-sm2}-\ref{eq:dn-sm2}) reduce to the standard LLE (e.g.~\cite{Coen2013, Hansson2013}) when TPA and free carriers are not present.

\begin{figure}[tbp]
\begin{center}
\includegraphics[width=1.00\columnwidth]{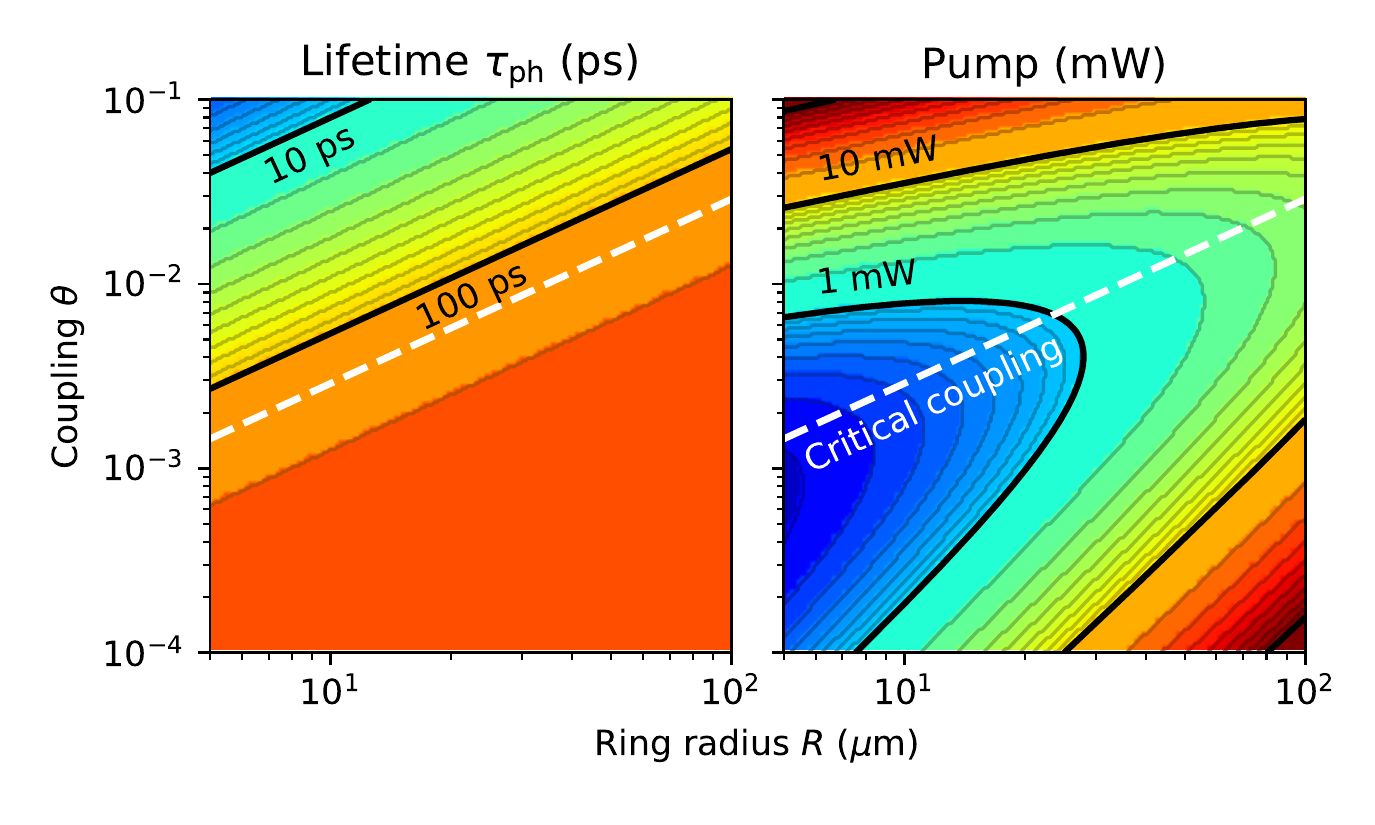}
\caption{Photon lifetime $\tau_{\rm ph}$ and pump power at $\bar{S} = 1$ for a microring cavity with coupled to a waveguide, as a function of ring radius and ring-waveguide coupling $\theta$ (power coupled per round trip).  $\mathcal{A}_{\rm eff} = 0.1\mu{\rm m}^2$, $\alpha = 2\,$dB/cm.  Based on Ref.~\cite{HamerlyMWP}}
\label{fig:mwp-f2}
\end{center}
\end{figure}

Fig.~\ref{fig:mwp-f2} plots the lifetime and scale of the nonlinearity (pump power required to set $\bar{S} = 1$) for a microring cavity with typical dimensions at 1.55$\mu$m (effective area $\mathcal{A}_{\rm eff} = 0.1\mu{\rm m}^2$ for $550\times 220$-nm ridge with 70-nm slab \cite{Koos2007}, loss $\alpha = 2.0\,$dB/cm).  Critically, pump powers are confined to a reasonable range $\lesssim 5$ mW for most structures, while photon lifetimes can vary in the range 10--200 ps.  Note that the combination of low loss and efficient carrier sweep-out makes the carrier lifetime shorter than the photon lifetime $\tau_c \lesssim \tau_{\rm ph}$, distinct from the condition $\tau_c \gg \tau_{\rm ph}$ typically encountered in integrated photonics.

\section{Gain and Phase Matching}
\label{sec:gain}

\begin{figure}[tbp]
\begin{center}
\includegraphics[width=1.00\columnwidth]{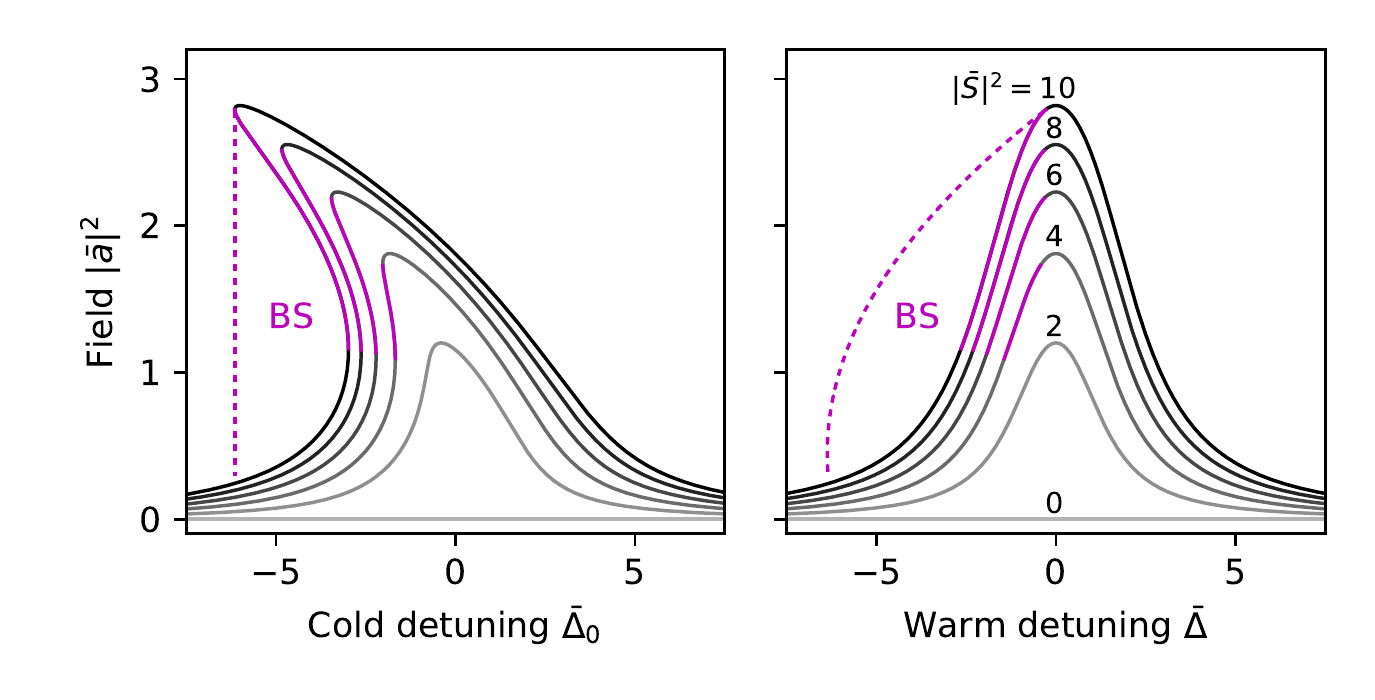}
\caption{Optical resonance and bistability, and difference between cold-cavity ($\bar{\Delta}_0$) and warm-cavity ($\bar{\Delta}$) detuning.  Unstable solutions due to optical bistability (BS) are shown in pink.  $\lambda = 1.55\mu$m, $\tau_c/\tau_{\rm ph} = 0.2$.}
\label{fig:f7}
\end{center}
\end{figure}

Next we consider the case of parametric amplification $2\omega_p \rightarrow \omega_s + \omega_i$: the cavity is pumped at resonance $\omega_p$, and four-wave mixing creates photon pairs at signal and idler frequencies $(\omega_s, \omega_i$).  The circulating field takes the form:
\beq
	\bar{a}(\bar{t},\bar{\tau}) = a_p(\bar{\tau}) + a_s(\bar{\tau})e^{ik\bar{t}} + a_i(\bar{\tau})e^{-ik\bar{t}} \label{eq:ft}
\eeq
It is straightforward to derive equations for ($a_p$, $a_s$, $a_i$) by applying (\ref{eq:ft}) to the Lugiato-Lefever equations (\ref{eq:da-sm2}-\ref{eq:dn-sm2}).  The derivation here is a simple extension of Ref.~\cite{Hansson2013} to the case with free carriers.  To assess whether parametric oscillation is possible, we first find the steady-state pump field by solving the bistability quintic
\beq
	|\bar{S}|^2 = \Bigl[\bigl(1 + r |a_p|^2 + \tfrac{\bar{\chi}_c}{\mu}|a_p|^4\bigr)^2 
		+ {\underbrace{\bigl(\bar{\Delta}_0 \!-\! |a_p|^2 \!+\! \bar{\chi}_c |a_p|^4\bigr)}_{\bar{\Delta}}}^2\Bigr]|a_p|^2
\eeq
for the steady-state $a_p$.  Fig.~\ref{fig:f7} shows typical bistability curves (note the curves bend to the left because the carrier index shift is negative) and the relation between cold- and warm-cavity detuning $\bar{\Delta} = \bar{\Delta}_0 - |a_p|^2 + \bar{\chi}_c |a_p|^4$, which we use for convenience in our figures since cavity resonance occurs at $\bar{\Delta} = 0$.  
Linearizing (\ref{eq:da-sm2}-\ref{eq:dn-sm2}) in terms of $(a_s, a_i)$ one determines stability.  The solutions take the exponential form $a_{s,i}(\bar{\tau}) = a_{s,i}(0) e^{g \bar{\tau}}$, where $g$ is an eigenvalue of the Jacobian:
\bea
	g & \!=\! & -1 - 2r|a_p|^2 - \mu^{-1}\bar{\chi}_c |a_p|^4 \nonumber \\
	& & \ \pm\, \sqrt{(1+r^2)|a_p|^4 - \bigl(\bar{\delta} - \eta k^2\bigr)^2} \label{eq:g-mm}
\eea
The parameter $\bar{\delta} \equiv \bar{\Delta} - |a_p|^2$ denotes the phase mismatch.  The gain in (\ref{eq:g-mm}) is maximized when the phase-matching condition $\eta k^2 = \bar{\delta}$ is satisfied.  The phase-matched case is treated below in Sec.~\ref{sec:pm-gain}, and the general case in Sec.~\ref{sec:mm-gain}.

\subsection{Phase-Matched Case}
\label{sec:pm-gain}

\begin{figure}[bp]
\begin{center}
\includegraphics[width=1.00\columnwidth]{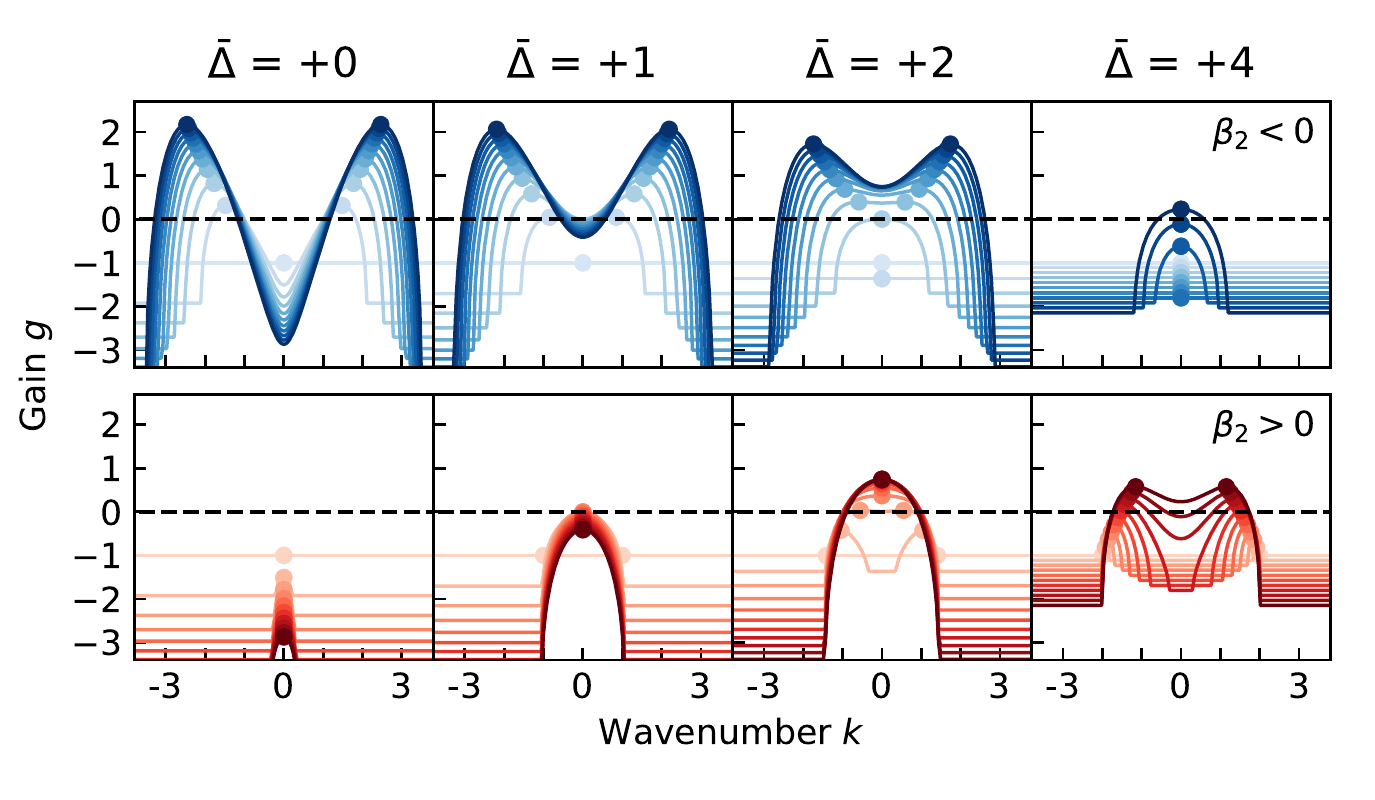}
\caption{Parametric gain $g(k)$ for various detuning, pump power, and dispersion values.  Pump power ranges from $|\bar{S}|^2 = 0$ (light) to $|\bar{S}|^2 = 50$ (dark).  $\tau_c/\tau_{\rm ph} = 0.1$.  Dots give $k$ at optimal gain in Eq.~(\ref{eq:kmax}).  Based on Ref.~\cite{HamerlyMWP}.}
\label{fig:f3}
\end{center}
\end{figure}

To achieve phase matching, $\bar{\delta}$ and $\eta$ must have the same sign.  For $\bar{\delta} > 0$ this requires normal dispersion, while for $\bar{\delta} < 0$ this requires anomalous dispersion.  Note that this differs from single-pass parametric amplification, where phase matching is only possible with anomalous dispersion.  Here it can occur for normal dispersion as well, thanks to the detuning of the cavity \cite{Hansson2013, Godey2014}.  With phase matching, the gain $g$ (called $g_{\rm max}$) is given by the formula:
\beq
	g_{\rm max} = -1 + (\sqrt{1 + r^2} - 2r)|a_p|^2 - \mu^{-1} \bar{\chi}_c |a_p|^4 \label{eq:gpm}
\eeq
This differs from the lossless case $g_{\rm max} = |a_p|^2 - 1$ \cite{Hansson2013} in two ways.  First, the $|a_p|^2$ term is reduced due to TPA.  The reduction depends on the ratio of real to imaginary nonlinear index $r = \beta\lambda/4\pi n_2$, which in turn is related to the nonlinear figure-of-merit \cite{Dinu2003}: $r = (4\pi F)^{-1}$.  Second, FCA leads to loss that scales with a higher power of $|a_p|$.  Unlike CaF$_2$, SiO$_2$ or Si$_3$N$_4$, where gain can always be achieved (in theory) by increasing pump power, this places a strict upper bound on gain in silicon.  Maximizing Eq.~(\ref{eq:gpm}) over $|a_p|^2$, substituting $\bar{\chi}_c$, and taking values for $1.55\mu$m from Appendix~\ref{sec:app-lle}, we see that the optimal gain is:
\bea
	g_{\rm max} & = & \frac{(\sqrt{1 + r^2} - 2r)^2 \hbar\omega \gamma v_g}{r \sigma} \frac{\tau_{\rm ph}}{\tau_c} - 1 \nonumber \\
	& = & 0.51 (\tau_{\rm ph}/\tau_c) - 1
	\label{eq:gmax-t}
\eea
This gives a strict condition for OPO, namely, gain will only occur if the ratio of carrier to photon lifetime is $\tau_c/\tau_{\rm ph} \lesssim 0.51$.  As discussed in Sec.~\ref{sec:mm}, this ratio is usually $\gg 1$ in silicon photonic structures, but can be reduced by controlling the carrier lifetime and simultaneously maintaining a low waveguide loss (high cavity $Q$).  With p-i-n carrier sweep-out (e.g.\ reducing the lifetime to $\leq 20\,{\rm ps}$ \cite{Gajda2011}, while maintaining losses of $\leq 2\,{\rm dB/cm}$), it is possible for this ratio to approach $\lesssim 0.1$ \cite{TurnerFoster2010, Rong2007}, tunable by the voltage applied to the diode \cite{Gajda2011}, allowing OPO to be realized in silicon at 1.55$\mu$m.

\begin{figure}[b!]
\begin{center}
\includegraphics[width=1.00\columnwidth]{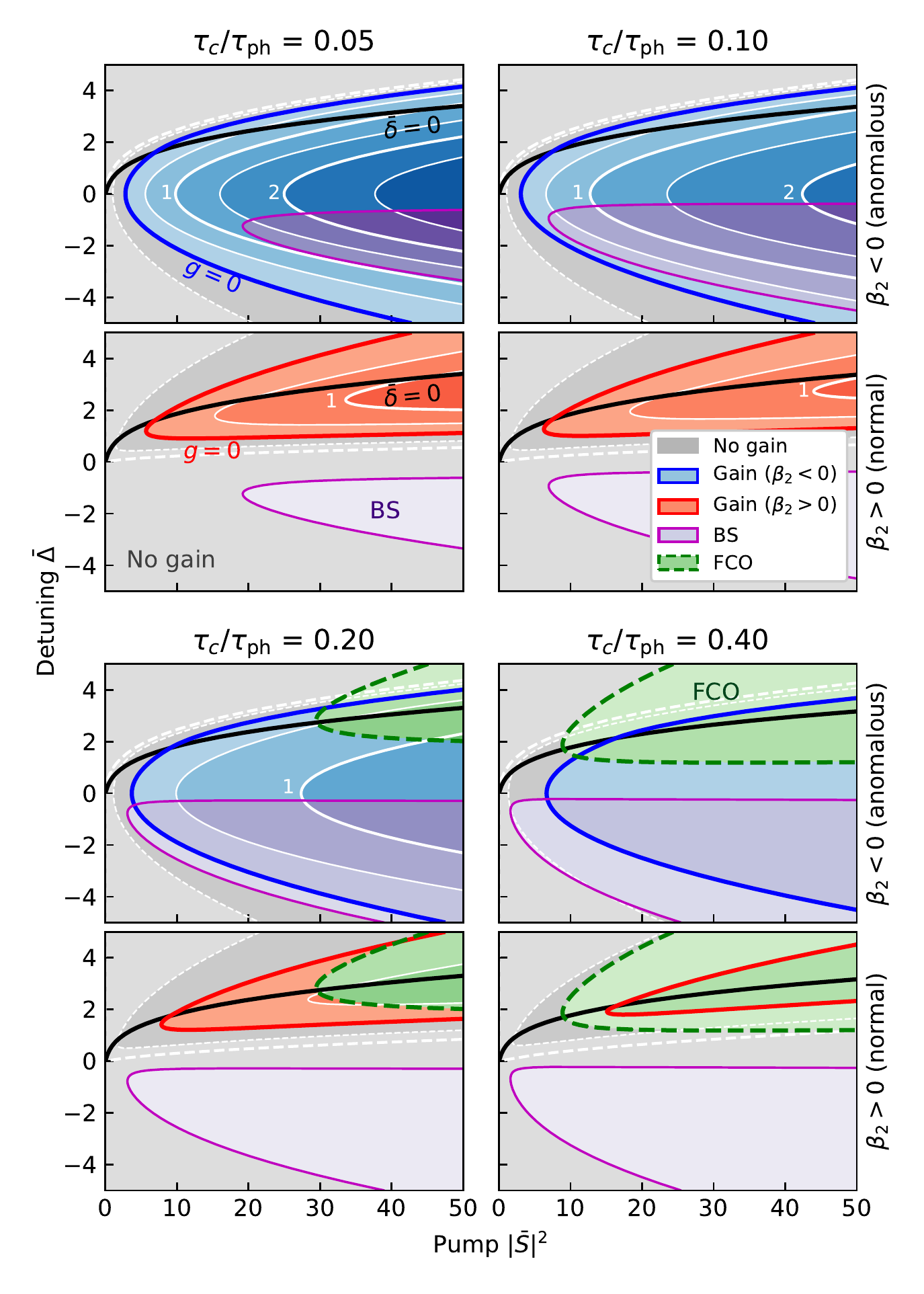}
\caption{Optimal gain $g(k_{\rm max})$ (white contours) for cavities with anomalous dispersion (blue) and normal dispersion (red), as a function of pump power and detuning.  Solid black line is gain threshold $g = 0$.  Black line denotes zero phase mismatch $\bar{\delta} = 0$.  Optical bistability (BS) and free-carrier oscillations (FCO) are given in the purple and green regions.  Based on Ref.~\cite{HamerlyMWP}.}
\label{fig:f2}
\end{center}
\end{figure}


\subsection{General Case}
\label{sec:mm-gain}

The gain spectrum $g(k)$ from Eq.~(\ref{eq:g-mm}) is plotted for a range of parameters ($\bar{\Delta}$, $\bar{S}$) in Fig.~\ref{fig:f3}.  The optimal gain, obtained by maximizing $g(k)$ over $k$, and depends on the dispersion:
\bea
	k_{\rm max} & \!\!=\!\! & \begin{dcases}
		\pm |\bar{\delta}|^{1/2} & \mbox{sgn}(\eta) = +\mbox{sgn}(\bar{\delta}) \\
		0 & \mbox{sgn}(\eta) = -\mbox{sgn}(\bar{\delta}) \end{dcases} \label{eq:kmax}
\eea
From Fig.~\ref{fig:f3}, we see that the gain profile is very dispersion-dependent.  As expected, near zero detuning, the gain spectrum is double-peaked for anomalous dispersion (as it is in single-pass waveguides \cite{Hansryd2002}) and the phase mismatch $\bar{\delta} > 0$ can be cancelled by dispersion.  For most parameters, the normal-dispersion gain is single-peaked suggesting phase matching is not satisfied except for sufficiently positive detunings.  For normal dispersion, a detuning is required to see gain at all, which raises the threshold power \cite{Haelterman1992}.

Figure \ref{fig:f2} shows the gain as a function of pump intensity $|\bar{S}|^2$ and warm-cavity detuning $\bar{\Delta}$.  Parametric oscillation occurs in the red and blue regions (normal and anomalous dispersion, respectively), $\bar{\delta} = 0$ is shown in black (normal GVD phase-matches above the curve, anomalous GVD phase-matches below it).  Although phase-matching helps one achieve gain, it is not necessary.  In particular, for $\tau_c/\tau_{\rm ph} \lesssim 0.20$ in Fig.~\ref{fig:f2}, a large part of the region $\bar{\delta} > 0$ experiences gain for normal GVD, even though phase matching is not satisfied.  This is important from an engineering standpoint, since the zero-dispersion wavelength for ridge waveguides as in Fig.~\ref{fig:mwp-f1}(b) is typically $\lambda_{\rm ZDM} \gtrsim 2\mu$m \cite{Yin2006}, and in many multi-project wafer photonics processes, the permitted ridge-waveguide dimensions do not allow for anomalous GVD.

Instabilities in the pump field can prevent parametric oscillation, and regions of instability are shown in Fig.~\ref{fig:f2}.  The solutions in the purple region are part of the unstable middle branch of the optical bistability (BS) curve (purple in Fig.~\ref{fig:f7}).  Free-carrier oscillations (FCO), a self-pulsing instability \cite{Malaguti2011} that occurs when $\tau_c \sim \tau_{\rm ph}$, occur in the green region.  These regions are determined by taking the $3\times 3$ Jacobian matrix $J$ of (\ref{eq:da-sm2}-\ref{eq:dn-sm2}) and searching for the conditions \cite{Hamerly2015}:
\beq
	\underbrace{\det(J) > 0}_{\rm BS},\ \ \ 
	\underbrace{(\mbox{tr}(J)^2-\mbox{tr}(J^2))\mbox{tr}(J)-2\det(J) > 0}_{\rm FCO}
\eeq
Parametric gain is dominant when the carrier lifetime is very small, while free-carrier oscillations take over for longer carrier lifetimes (comparable to the photon lifetime).  

The gain profiles in Fig.~\ref{fig:f3} are a function of wavenumber $k$.  Here $k$ and fast time $\bar{t}$ were normalized to set the GVD term to $\eta = \pm 1$: converting back to unnormalized units one has:
\bea
	t = \sqrt{|\beta_2| v_g \tau_{\rm ph}}\,\bar{t},\ \ \ 
	\Delta\omega = (|\beta_2| v_g \tau_{\rm ph})^{-1/2} k
\eea
Since the gain region tends to live around $|k| \lesssim 2$, this provides and order-of-magnitude estimate of the gain-bandwidth, primary-comb line spacing, and soliton bandwidth \cite{Coen2013} (in soliton regime):
\bea
	|\Delta\omega|_{\rm comb} & \lesssim & 2/\sqrt{|\beta_2|v_g\tau_{\rm ph}} \nonumber \\
	& \approx & 2\pi\times (5\,\mbox{THz})\left(\frac{|\beta_2|}{0.1{\rm ps}^2/{\rm m}} \frac{\tau_{\rm ph}}{0.5{\rm ns}}\right)^{-1/2} \label{eq:dw-comb}
\eea
which suggests that, for reasonable dispersion engineering \cite{Okawachi2014}, the comb bandwidth could be $\gtrsim 10\,$THz and pulses could be around $\lesssim\,0.1{\rm ps}$ long.

\section{Numerical Simulation}
\label{sec:sim}

\begin{figure}[tbp]
\begin{center}
\includegraphics[width=1.00\columnwidth]{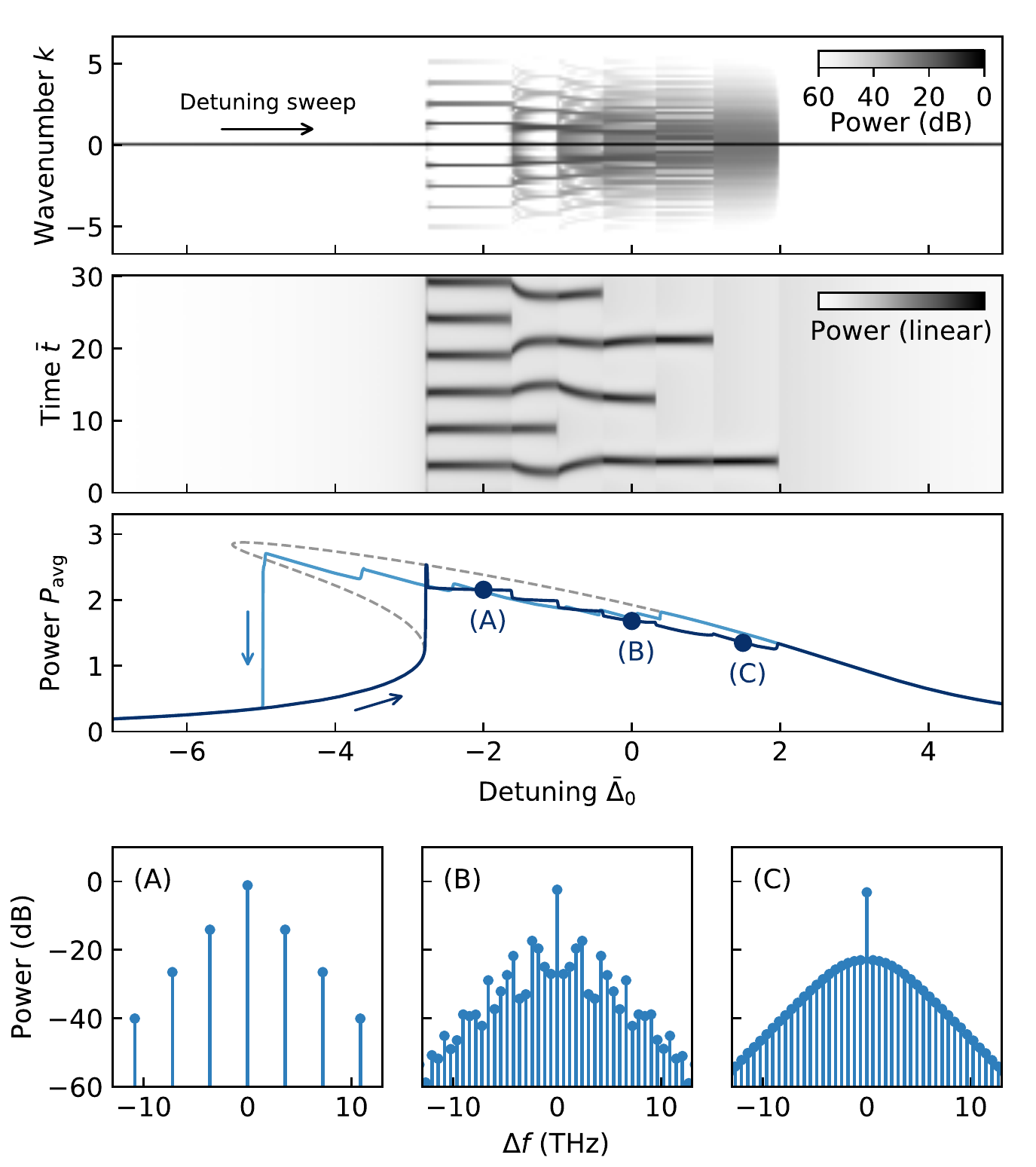}
\caption{LLE simulation of detuning sweep with anomalous dispersion.  Top: power spectrum and temporal shape of circulating field as detuning is increased.  Middle: average power in cavity for forward and reverse sweeps.  Bottom: spectra of Turing rolls, multi-soliton and single-soliton states (A-C).  $R=20\mu$m, $|\bar{S}|^2 = 10$, $\beta_2 = -0.2{\rm ps}^2/{\rm m}$, $\tau_{\rm ph} = 0.2{\rm ns}$, $\tau_c/\tau_{\rm ph} = 0.2$.}
\label{fig:f13}
\end{center}
\end{figure}

To provide an example, we simulate Eqs.~(\ref{eq:da-sm2}-\ref{eq:dn-sm2}) on two model systems with anomalous and normal dispersion, respectively.  The anomalous-dispersion case is shown in Fig.~\ref{fig:f13}.  The familiar progression from Turing rolls (A) to multi-soliton states (B) to a single soliton (C) is observed \cite{Godey2014, Herr2014}, but without an intervening chaotic state (possibly because the pump power is not sufficiently high \cite{Godey2014}).  Each transition is accompanied by a jump in power $P_{\rm avg} = t_R^{-1}\int{|\bar{a}(t)|^2\d t}$ \cite{Herr2014}.  For the parameters used here ($\beta_2 = -0.2{\rm ps}^2/{\rm m}$, $\tau_{\rm ph} = 0.2{\rm ns}$), Eq.~(\ref{eq:dw-comb}) predicts a bandwidth of 4.4 THz.  The primary comb (A) has a line spacing of 6 FSR in this case (3.6 THz) and the single-soliton spectrum (C) can be fit to sech$^2$ profile with $\Delta f_{\rm FWHM} = 5.2\,{\rm THz}$, in agreement with Eq.~(\ref{eq:dw-comb}).

\begin{figure}[tbp]
\begin{center}
\includegraphics[width=1.00\columnwidth]{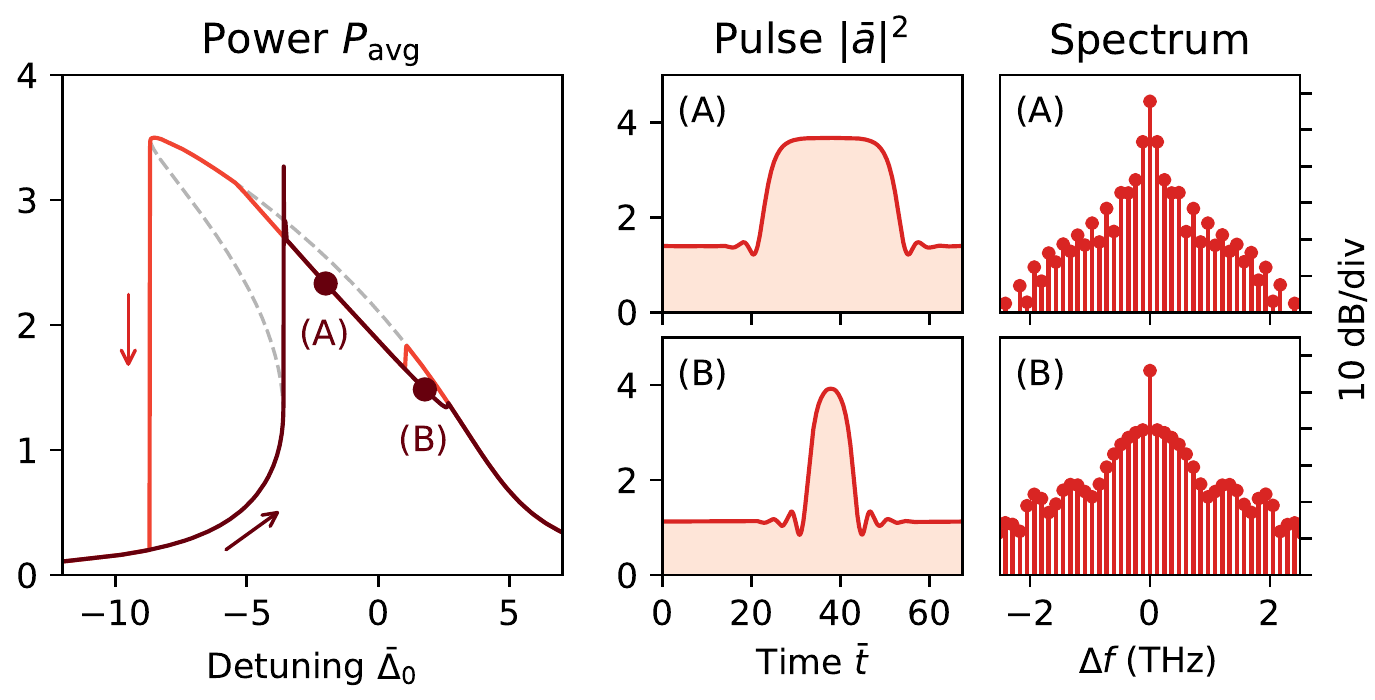}
\caption{Normal-dispersion LLE simulation.  Detuning is swept as in Fig.~\ref{fig:f13}.  Left: average power as function of detuning.  Right: pulse shape and spectrum of dark soliton at two different detuning values.  $R = 60\mu$m, $|\bar{S}|^2 = 16$, $\beta_2 = +1.0{\rm ps}^2/{\rm m}$, $\tau_{\rm ph} = 0.2{\rm ns}$, $\tau_c/\tau_{\rm ph} = 0.2$.}
\label{fig:f14}
\end{center}
\end{figure}

Fig.~\ref{fig:f14} shows a simulation for normal cavity dispersion.  Here, modulation instability leads to the formation of dark solitons \cite{Godey2014, Xue2015}.  These dark solitons form from the interaction of two domain walls connecting the lower and upper optical bistability curves of the cavity.  The spacing between the domain walls is a function of cavity detuning.

Free carriers cause some qualitative changes relative to combs in the lossless case (e.g.\ Si$_3$N$_4$).  Note that the bistability curves in Figs.~\ref{fig:f13}-\ref{fig:f14} bend to the left rather than right, since the carrier-induced dispersion is larger than the Kerr effect in these cases.  This gives rise to a sudden turn-on of parametric oscillation when the system moves from the lower to upper bistability branch.  $P_{\rm avg}$ in the soliton regime also decreases with $\bar{\Delta}_0$ (contrary to \cite{Herr2014}), which may complicate thermal locking or require active stabilization for these particular cases.


\section{Wavelength Dependence}
\label{sec:lamdep}

As the operating wavelength is increased $n_2$ grows while two-photon absorption $\beta$ is either reduced or eliminated; consequently the conditions for parametric amplification are relaxed.  This is why amplification can be observed at $\gtrsim 2.0\mu$m in ordinary structures without p-i-n sweepout \cite{Liu2010}.  This section derives the general conditions in which amplification is possible, in terms of wavelength $\lambda$, waveguide loss $\alpha$ and carrier lifetime $\tau_c$.

\subsection{Limit on Carrier Lifetime}

\begin{figure}[tbp]
\begin{center}
\includegraphics[width=1.0\columnwidth]{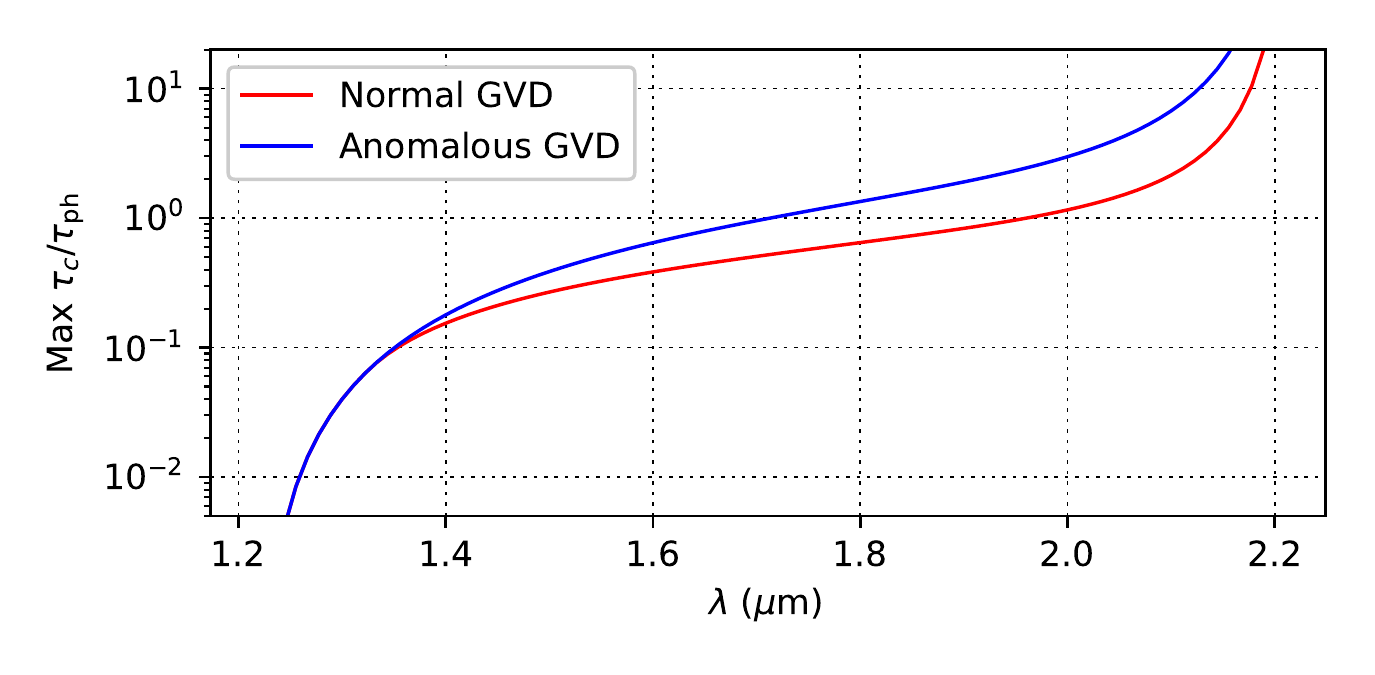}
\caption{Carrier-lifetime limit $(\tau_c/\tau_{\rm ph})_{\rm max}$ for anomalous and normal GVD.}
\label{fig:f8}
\end{center}
\end{figure}

Eq.~(\ref{eq:gmax-t}) sets the maximum carrier lifetime (relative to photon lifetime) for which net gain is possible:
\beq
	(\tau_c/\tau_{\rm ph})_{\rm max} = \frac{(\sqrt{1 + r^2} - 2r)^2 \hbar\omega \gamma v_g}{r \sigma} \label{eq:taumax}
\eeq
This quantity (blue curve in Fig.~\ref{fig:f8}) depends on wavelength through $\omega$ and material constants $r, \gamma, \sigma$.  This limit is applicable in the anomalous-dispersion regime, where an oscillation region always exists around $\bar{\Delta} = 0$ if $\tau_c/\tau_{\rm ph} < (\tau_c/\tau_{\rm ph})_{\rm max}$ (see Fig.~\ref{fig:f2}).

In the normal-dispersion case there is a more stringent bound set by free-carrier oscillations.  Fig.~\ref{fig:f2} shows that for $\tau_{c}/\tau_{\rm ph} \geq 0.40$, it is impossible to reach the (red) parametric-gain region without first crossing into the (green) FCO region.  Once the free-carrier oscillations start, they disrupt the parametric process and suppress gain.  The FCO region overtakes the gain region roughly when the red/blue ($g_{\rm max} = 0$), black ($\bar{\delta} = 0$) and green lines in Fig.~\ref{fig:f2} intersect at the same point, calculated in Fig.~\ref{fig:f8} (red curve).

These conditions can be recast in terms of the effective loss $\alpha' = \alpha + \theta/L$, since $\tau_{\rm ph} = 1/(\alpha' v_g)$.  Thus oscillation places a constraint on the figure-of-merit $F = 1/\sqrt{\tau_c \alpha'}$ of Ref.~\cite{TurnerFoster2010}.  Long carrier lifetimes can be alleviated with very low-loss structures; likewise high loss can be compensated with short carrier lifetimes.

\subsection{Limit on Waveguide Loss}

\begin{figure}[tbp]
\begin{center}
\includegraphics[width=1.0\columnwidth]{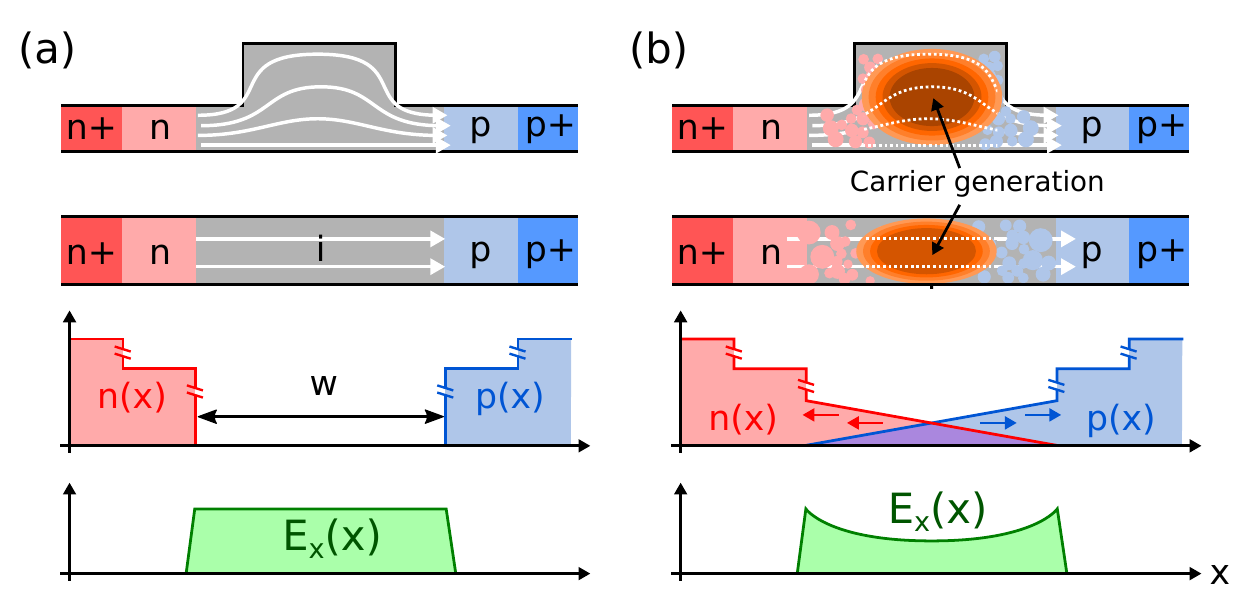}
\caption{Illustration of free-carrier sweep-out field screening.  State of waveguide with (a) weak field and (b) strong field generating carriers.  Top to bottom: 2D model of waveguide.  Approximate 1D model.  Steady-state carrier densities $n(x), p(x)$ in 1D model.  Steady-state electric field $E_x(x)$ in 1D model.}
\label{fig:f10}
\end{center}
\end{figure}

We can derive an upper bound on the effective loss $\alpha'$ that arises from screening of the free-carrier sweepout field.  Screening limits the field intensity $I$, and since the intensity at the OPO threshold is inversely proportional to cavity loss, this places a limit on $\alpha'$ for which gain can be observed.  Fig.~\ref{fig:f10} illustrates screening for a 2D waveguide: at large field intensities, since the extraction process is not instantaneous, the steady-state carrier distribution effectively screens the reverse-bias field \cite{Dimitropoulos2005}.  If this screening reduces the internal field to near zero, carrier sweep-out will be ineffective.

A full treatment of screening requires sophisticated 2D transport simulations \cite{Gajda2011}; however, a reasonable analytic estimate can be derived by approximating the waveguide as a 1D system.  The transport equations become:
\bea
	\frac{\partial p}{\partial t} & = & G(x) - \frac{\partial}{\partial x} (v_{p}(E_x) p) \\
	\frac{\partial n}{\partial t} & = & G(x) + \frac{\partial}{\partial x} (v_{n}(E_x) n) \\
	\frac{\partial E_x}{\partial x} & = & \frac{e}{\epsilon_0} (p - n)
\eea
To minimize the carrier lifetime, p-i-n structures are strongly reverse-biased, so the electron and hole drift velocities are close to the saturation velocity $v_{\rm sat} \approx 10^7 {\rm cm}/{\rm s}$ \cite{Jacoboni1977}.  Assuming a uniform carrier generation rate $G$, the steady-state carrier concentrations are $p = G x/v_{\rm sat}$, $n = G(w-x)/v_{\rm sat}$, where $w$ is the spacing between the doped regions.  Further assuming that the voltage drop across the doped regions is small compared to the drop across the intrinsic region (so $\int_0^w{E_x\d x} = V_0$), we find:
\beq
	E_x = \left(\frac{V_0}{w} - \frac{e G w^2}{12\epsilon_0 v_{\rm sat}}\right) + \frac{e G}{\epsilon_0 v_{\rm sat}} (x - w/2)^2
\eeq
This field is minimized at the center of the waveguide.  Screening will prevent carrier sweep-out when $E_x(w/2) \approx 0$, which limits the generation rate to:
\beq
	G \lesssim \frac{12\epsilon_0 v_{\rm sat} V_0}{e w^3} \label{eq:maxgen}
\eeq
For typical p-i-n parameters ($w = 1\mu$m, $V_0 = 15\,$V), one finds $G \lesssim 10^{27}{\rm cm}^{-3}{\rm s}^{-1}$, which corresponds to $I \lesssim 4\times 10^{8}{\rm W}/{\rm cm}^2$ at 1.55$\mu$m (about $400\,$mW for $A_{\rm eff} = 0.1\,\mu{\rm m}^2$).  Numerical simulations on waveguides of this size confirm this bound \cite{Gajda2011}.

%

Gain depends only on the circulating field through Eq.~(\ref{eq:gpm}).  The gain threshold intensity, scaled to unnormalized units (Appendix~\ref{sec:app-lle}) and setting $\zeta \equiv \tau_c/\tau_{c,\rm max}$, is:
\beq
	I = |a|^2 = \frac{\alpha'}{\gamma(\sqrt{1 + r^2} - 2r)} \underbrace{\frac{1 - \sqrt{1 - \zeta}}{\zeta}}_{f(\zeta)}
\eeq
where the $\zeta$-dependent term limits to $f(\zeta) = \tfrac{1}{2}$ for $\zeta \ll 1$ ($\tau_c \ll \tau_{c,\rm max}$) and $f(\zeta) = 1$ at $\zeta = 1$ ($\tau_c = \tau_{c,\rm max}$).  Since the carrier generation rate for two-photon absorption is $G = \beta I^2/2\hbar\omega = (r\gamma/\hbar\omega) I^2$, one can bound $\alpha'$ using Eq.~(\ref{eq:maxgen}):
\beq
	\alpha' \lesssim \frac{\sqrt{1+r^2}-2r}{f(\zeta)} \left[\frac{12\epsilon_0 \hbar\omega \gamma v_{\rm sat} V_0}{r e w^3}\right]^{1/2} \label{eq:alphamax}
\eeq
This bounds effective loss per unit length due to both input/output coupling and intrinsic roughness / absorption.  Equivalently, it is a bound on the resonator's loaded $Q_L = 2\pi n_g/\lambda\alpha'$.  Thus, the bound on intrinsic $\alpha$ becomes stricter the more over-coupled a cavity is.  At $2\mu$m, the bound is very loose (tens of dB/cm), but at telecom wavelengths is of order a few dB/cm, depending on the cavity coupling.

\begin{figure}[tbp]
\begin{center}
\includegraphics[width=1.0\columnwidth]{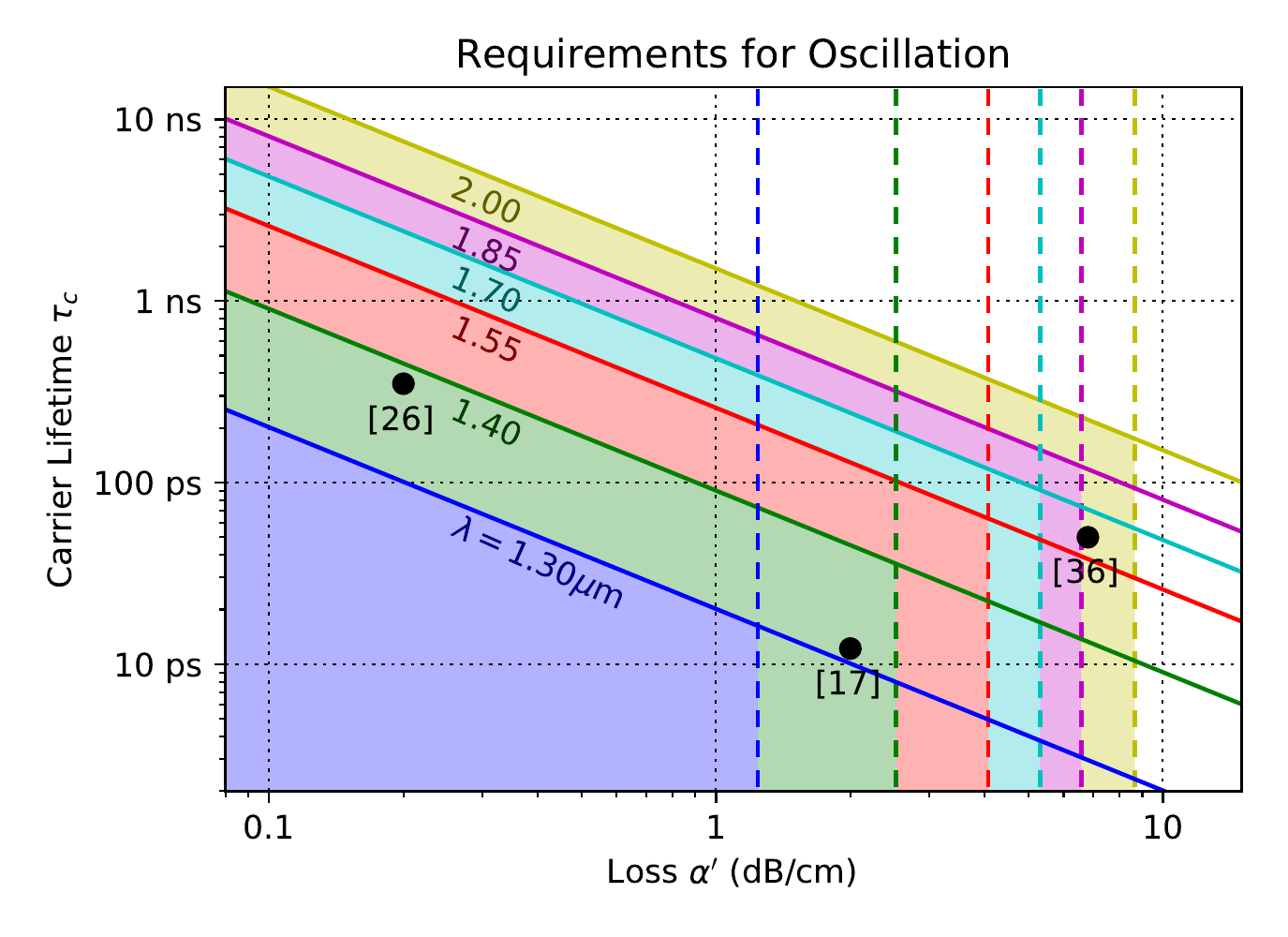}
\caption{Silicon ring-resonator conditions for parametric oscillation.  Solid lines denote limit Eq.~(\ref{eq:taumax}), while dashed lines denote limit Eq.~(\ref{eq:alphamax}) ($f(\zeta) = 1$, $V_0 = -15\,{\rm V}$, $w = \lambda/1.55$).}
\label{fig:f11}
\end{center}
\end{figure}

Conditions (\ref{eq:taumax}) and (\ref{eq:alphamax}) are plotted together in Fig.~\ref{fig:f11}.  This provides a comprehensive picture of the design requirements for parametric oscillation in silicon ring cavities with active carrier removal.  While the requirements for long-wavelength oscillation are relatively loose, oscillation at or below 1.55$\mu$m require both low-loss structures ($\lesssim 2$--$3\,{\rm dB/cm}$) with short carrier lifetimes ($\lesssim 100\,{\rm ps}$), which will require careful engineering of the fabrication process and resonator design.

\section{Conclusion}

We have studied the conditions for parametric oscillation in silicon microring cavities, focusing primarily on phenomena at 1.55$\mu$m, but with an analysis that generalizes to all wavelengths.  Parametric oscillation allows the formation of frequency combs in both the anomalous- and normal-dispersion regimes.  The key advances that make this possible are the fabrication of ridge waveguides with losses $\lesssim 2$--$3\,$dB/cm, and carrier sweep-out using a p-i-n junction, which can reduce carrier lifetimes to $\lesssim 10\,$ps while maintaining low propagation losses.  With typical silicon waveguide dimensions, it should be possible to form bright- and dark-soliton combs at $1.55\mu$m with bandwidths $\gtrsim\,$THz.  Proper dispersion engineering could enable broader combs.

Extensions of this work could treat Raman scattering and thermal effects, which were not considered in this paper.  In addition, while this paper showed numerically that frequency combs are possible in principle, a more thorough study will be required to determine precisely how combs in the free-carrier regime differ from their conventional counterparts.  Our model may also be applicable to other material platforms where multiphoton absorption and photogenerated carriers are present and limit performance.  




\appendices
\section{Parameters in Normalized LLE}
\label{sec:app-lle}

\begin{table}[bp]
\renewcommand{\arraystretch}{1.3}
\caption{Constants Used in Normalized Lugiato-Lefever Equation}
\label{tab:t1}
\centering
\begin{tabular}{rlc|c}
\hline\hline
\multicolumn{4}{c}{Normalization Constants} \\ \hline
$\xi_a=$ & $(2\gamma v_g \tau_{\rm ph})^{-1/2}$ & & Circulating field \\
$\xi_n=$ & $1/\mu \sigma v_g \tau_{\rm ph}$ & & Carrier density \\
$\xi_{\rm in}=$ & $\sqrt{t_R^2/8\gamma v_g \tau_{\rm ph}^3\theta}$ & & Input/output field \\
$\xi_t=$ & $\sqrt{|\beta_2| v_g \tau_{\rm ph}}$ & & Fast time \\
\hline
\multicolumn{4}{c}{Material Constants and Values at 1.55$\mu$m} \\ \hline
$r=$ & $\beta\lambda /4\pi n_2$ & 0.189 & TPA / Kerr ratio \\
$\gamma=$ & $2\pi n_2/\lambda$ & $3.1\!\cdot\! 10^{-9}{\rm cm}/{\rm W}$ & Nonlinear refraction \\
$\mu=$ & $\tfrac{4\pi}{\lambda} \frac{|\d n/\d n_c|}{\d\alpha/\d n_c}$ & 25 & FCD / FCA ratio \\
$\sigma=$ & $\d\alpha/\d n_c$ & $1.45\!\cdot\! 10^{-17}{\rm cm}^2$ & FCA cross section \\
\hline\hline
\end{tabular}
\end{table}

\begin{figure}[!b]
\begin{center}
\includegraphics[width=1.0\columnwidth]{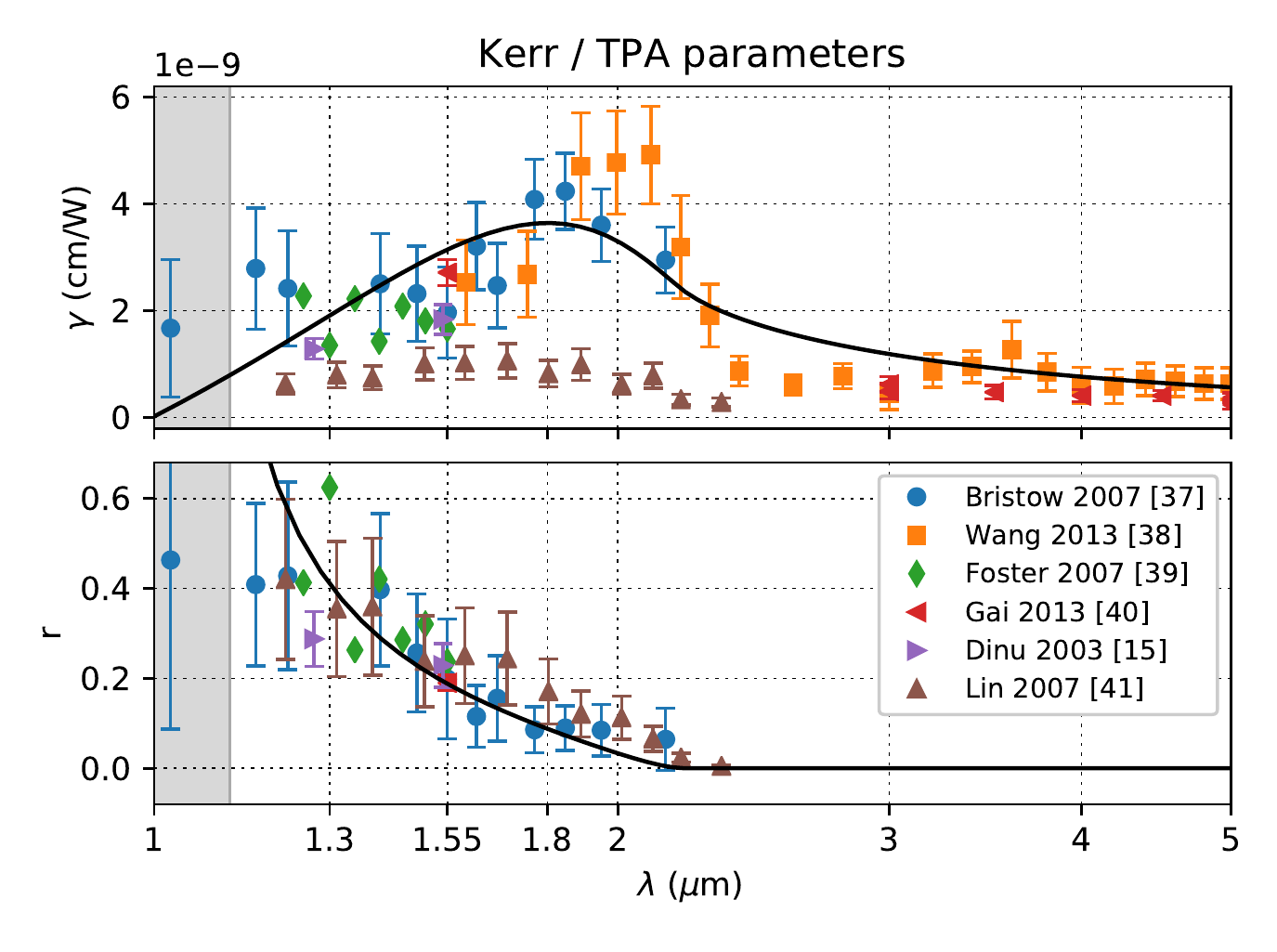}
\caption{Scaling of Kerr and TPA parameters $\gamma$ and $r$ with wavelength.  Solid lines are fit using indirect band-gap parabolic-band models \cite{Garcia2006, Hon2011}.}
\label{fig:f9}
\end{center}
\end{figure}

The terms in the Lugiato-Lefever Equation (LLE) are normalized according to: $a = \xi_a \bar{a}$, $n_c = \xi_n \bar{n}_c$, $a_{\rm in} = \xi_{\rm in} \bar{S}$, $t = \xi_t \bar{t}$, $\tau = \tau_{\rm ph} \bar{\tau}$.  Here $|a|^2$ and $|a_{\rm in}|^2$ are intensities (units W/cm$^2$) and $n_c$ is a carrier density (units cm$^{-3}$).  The normalization constants are given in Table~\ref{tab:t1}.  For the group velocity $v_g = c/n_g$, the value for silicon $n_g = 3.6$ was used.

The material constants in Table~\ref{tab:t1} depend on the nonlinear index $n_2$, two-photon absorption $\beta$, and free-carrier index change and absorption $\d n/\d n_c$, $\d\alpha/\d n_c$ (electrons plus holes), which are tabulated in the literature.  Both $\gamma$ and $r$ depend nontrivially on $\lambda$ (Fig.~\ref{fig:f9}), and $\mu \propto \lambda^{-1}$ and $\sigma \propto \lambda^2$ follow Drude-model scaling.  The photon lifetime is given by $\tau_{\rm ph} = 1/\alpha' v_g$, where $\alpha' = \alpha + \theta / L$ is the effective loss per unit length (intrinsic plus coupling).


\section*{Acknowledgment}

The authors thank Meysam Namdari and Levon Mirzoyan for useful discussions.  This work is supported (in part) by the German Research Foundation (DFG) within the project: Silicon-on-Insulator based Integrated Optical Frequency Combs.  R.H.\ was supported by the ImPACT Program of the Council of Science, Technology and Innovation (Cabinet Office, Government of Japan) and an appointment to the IC Postdoctoral Research Fellowship Program at MIT, administered by ORISE through U.S.\ DOE / ODNI.

\ifCLASSOPTIONcaptionsoff
  \newpage
\fi


\begin{thebibliography}{1}

\bibitem{Pasquazi2018}{A.~Pasquazi, M.~Peccianti, L.~Razzari, D.~J.~Moss, S.~Coen et al., ``Micro-combs: A novel generation of optical sources.'' {\it Phys.\ Rep.}, vol.~729, pp.~1--81, 2018.}
\bibitem{Inagaki2016}{T.~Inagaki, K.~Inaba, R.~Hamerly, K.~Inoue, Y.~Yamamoto, and H.~Takesue, ``Large-scale Ising spin network based on degenerate optical parametric oscillators.'' {\it Nat.\ Photonics}, vol.~10, pp.~415--419, 2016.}
\bibitem{Roslund2014}{J.~Roslund, R.~M.~de~Adra\'{u}jo, S.~Jiang, C.~Fabre, and N.~Treps, ``Wavelength-multiplexed quantum networks with ultrafast frequency combs.'' {\it Nat.\ Photonics}, vol.~8, pp.~109--112, 2013.}
\bibitem{Pfeifle2014}{J.~Pfeifle, V.~Brasch, M.~Lauermann, Y.~Yu et al., ``Coherent terabit communications with microresonator Kerr frequency combs.'' {\it Nat.\ Photonics} vol.~8, no.~5, pp.~375--380, 2014.}
\bibitem{Kitayama1997}{K.~Kitayama, ``Highly stabilized millimeter-wave generation by using fiber-optic frequency-tunable comb generator.'' {\it J.\ Lightw.\ Technol.}, vol.~15, no.~5, pp.~883--893, 1997.}
\bibitem{Kippenberg2004}{T.~Kippenberg, S.~Spillane, and K.~Vahala, ``Kerr-nonlinearity optical parametric oscillation in an ultrahigh-Q toroid microcavity.'' {\it Phys.~Rev.~Lett.} vol.~93, no.~083904, 2004}
\bibitem{Herr2014}{T.~Herr, V.~Brasch, J.~D.~Jost, C.~Y.~Wang, N.~M.~Kondratiev, M.~L.~Gorodetsky, and T.~J.~Kippenberg, ``Temporal solitons in optical microresonators.'' {\it Nat.\ Photonics} vol.~8, pp.~145--152, 2014.}
\bibitem{Chembo2010}{Y.~K.~Chembo, D.~V.~Strekalov, and N.~Yu, ``Spectrum and Dynamics of Optical Frequency Combs Generated with Monolithic Whispering Gallery Mode Resonators.'' {\it Phys.~Rev.~Lett.} vol.~104, no.~103902, 2010.}
\bibitem{Saha2013}{K.~Saha, Y.~Okawachi, B.~Shim, J.~S.~Levy, R.~Salem et al., ``Modelocking and femtosecond pulse generation in chip-based frequency combs.'' {\it Opt.~Express} vol.~21, pp.~1335--1343, 2013.}
\bibitem{Pfeiffer2016}{M.~H.~P.~Pfeiffer, A.~Kordts, V.~Brasch, M.~Zervas, ``Photonic Damascene process for integrated high-Q microresonator based nonlinear photonics.'' {\it Optica} vol.~3, no.~1, pp.~20--25, 2016.}
\bibitem{Foster2006}{M.~A.~Foster, A.~C.~Turner, J.~E.~Sharping, B.~S.~Schmidt, M.~Lipson, and A.~L.~Gaeta, ``Broad-band optical parametric gain on a silicon photonic chip.'' {\it Nature} vol.~441, pp.~960--963, 2006.}
\bibitem{HamerlyMWP}{R.~Hamerly, M.~Namdari, L.~Mirzoyan, D.~Gray, C.~Rogers and K.~Jamshidi, ``Conditions for Parametric and Free-Carrier Oscillation in SOI Ring Cavities with Active Carrier Removal,'' presented at Microwave Photonics (MWP) 2017, Beijing, China, Oct 23-26, 2017.}
\bibitem{Liu2010}{X.~Liu, R.~M.~Osgood~Jr., Y.~A.~Vlasov, and W.~M.~J.~Green, ``Mid-infrared optical parametric amplifier using silicon nanophotonic waveguides,'' {\it Nat.\ Photonics} vol.~4, pp.~557-560, 2010.}
\bibitem{Griffith2015}{A.~G.~Griffith, R.~K.~W.~Lau, J.~Cardenas, Y.~Okawachi, A.~Mohanty et al., ``Silicon-chip mid-infrared frequency comb generation.'' {\it Nat.~Comm.} vol.~6, no.~6299, 2015.}
\bibitem{Dinu2003}{M.~Dinu, F.~Quochi, and H.~Garcia, ``Third-order nonlinearities in silicon at telecom wavelengths.'' {\it Appl.\ Phys.\ Lett.} vol.~82, no.~18, pp.~2954--2956, 2003.}
\bibitem{Dekker2007}{R.~Dekker, N.~Usechak, M.~Forst, and A.~Driessen, ``Ultrafast nonlinear all-optical processes in silicon-on-insulator waveguides.'' {\it J.\ Phys.\ D: Appl.\ Phys.} vol.~40, pp.~R249--R271, 2007.}
\bibitem{TurnerFoster2010}{A.~C.~Turner-Foster, M.~A.~Foster, J.~S.~Levy, C.~B.~Poitras et al., ``Ultrashort free-carrier lifetime in low-loss silicon nanowaveguides.'' {\it Opt.\ Express} vol.~18, no.~4, pp.~3582-3591, 2010.}
\bibitem{Gajda2011}{A.~Gajda, L.~Zimmermann, J.~Bruns, B.~Tillack, and K.~Petermann, ``Design rules for p-i-n diode carriers sweeping in nano-rib waveguides on SOI.'' {\it Opt.\ Express} vol.~19, no.~10, pp.~9915--9922, 2011.}
\bibitem{Lugiato1987}{L.~A.~Lugiato and R.~Lefever, ``Spatial dissipative structures in passive optical systems,'' {\it Phys.\ Rev.\ Lett.} vol.~58, no.~21, pp.~2209--2211, 1987.}
\bibitem{Hamerly2017}{R.~Hamerly, D.~Gray, C.~Rogers, L.~Mirzoyan, M.~Namdari and K.~Jamshidi, ``Optical bistability, self-pulsing and XY optimization in silicon micro-rings with active carrier removal.'' in {\it Proc.\ of SPIE}, San Francisco, CA, 2017, pp.~10098-11}
\bibitem{Lin2007}{Q.~Lin, O.~J.~Painter, and G.~P.~Agrawal, ``Nonlinear optical phenomena in silicon waveguides: Modeling and applications.'' {\it Opt.\ Express} vol.~15, no.~25, pp.~16604--16644, 2007.}
\bibitem{Coen2013}{S.~Coen and M.~Erkintalo, ``Universal scaling laws of Kerr frequency combs.'' {\it Opt.\ Letters} vol.~38, no.~11, pp.~1790--1792, 2013.}
\bibitem{Hansson2013}{T.~Hansson, D.~Modotto, and S.~Wabnitz, ``Dynamics of the modulational instability in microresonator frequency combs.'' {\it Phys.\ Rev.\ A} vol.~88, no.~023819, 2013.}
\bibitem{Koos2007}{C.~Koos, L.~Jacome, C.~Poulton, J.~Leuthold, and W.~Freude, ``Nonlinear silicon-on-insulator waveguides for all-optical signal processing.'' {\it Opt.\ Express} vol.~15, no.~10, pp.~5976--5990, 2007.}
\bibitem{Godey2014}{C.~Godey, I.~V.~Balakireva, A.~Coillet, and Y.~K.~Chembo, ``Stability analysis of the spatiotemporal Lugiato-Lefever model for Kerr optical frequency combs in the anomalous and normal dispersion regimes.'' {\it Phys.\ Rev.\ A} vol.~89, no.~063814, 2014.}
\bibitem{Rong2007}{H.~Rong, S.~Xu, Y-H.~Kuo, V.~Sih et al., ``Low-threshold continuous-wave Raman silicon laser.'' {\it Nat.\ Photonics} vol.~1, pp.~232--237, 2007.}
\bibitem{Hansryd2002}{J.~Hansryd, P.~Andrekson, M.~Westlund, J.~Li, and P.~Hedekvist, ``Fiber-based optical parametric amplifiers and their applications.'' {\it IEEE J.\ Sel.\ Topics Quantum Electron.}, vol.~8, no.~3, pp.~506--520, 2002.}
\bibitem{Haelterman1992}{M.~Haelterman, S.~Trillo, and S.~Wabnitz, ``Dissipative modulation instability in a nonlinear dispersive ring cavity.'' {\it Opt.~Comm.} vol.~91, pp.~401--407, 1992.}
\bibitem{Yin2006}{L.~Yin, Q.~Lin, and G.~P.~Agrawal, ``Dispersion tailoring and soliton propagation in silicon waveguides,'' {\it Opt.\ Lett.} vol.~31, no.~9, pp.~1295-1297, 2006.}
\bibitem{Malaguti2011}{S.~Malaguti, G.~Bellanca, A.~de~Rossi, S.~Combri\'{e}, and Stefano~Trillo, ``Self-pulsing driven by two-photon absorption in semiconductor nanocavities,'' {\it Phys.\ Rev.\ A} vol.~83, no.~051802(R), 2011.}
\bibitem{Hamerly2015}{R.~Hamerly and H.~Mabuchi, ``Optical Devices Based on Limit Cycles and Amplification in Semiconductor Optical Cavities,'' {\it Phys.\ Rev.\ Appl.} vol.~4, no.~024016, 2015.}
\bibitem{Okawachi2014}{Y.~Okawachi, M.~R.~E.~Lamont, K.~Luke, D.~O.~Carvalho et al., ``Bandwidth shaping of microresonator-based frequency combs via dispersion engineering.'' {\it Opt.\ Letters} vol.~39, no.~12, pp.~3535--3538, 2014.}
\bibitem{Xue2015}{X.~Xue, Y.~Xuan, Y.~Liu, P-H.~Wang. S.~Chen et al., ``Mode-locked dark pulse Kerr combs in normal-dispersion microresonators.'' {\it Nat.\ Photonics} vol.~9, pp.~594--600, 2015.}
\bibitem{Dimitropoulos2005}{D.~Dimitropoulos, S.~Fathpour, and B.~Jalali, ``Limitations of active carrier removal in silicon Raman amplifiers and lasers,'' {\it Appl.\ Phys.\ Lett.} vol.~87, no.~261108, 2005.}
\bibitem{Jacoboni1977}{C.~Jacoboni, C.~Canali, G.~Ottaviani, and A.~A.~Quaranta, ``A review of some charge transport properties of silicon.'' {\it Solid State Electron.} vol.~20, no.~2, pp.~77-89, 1977.}
\bibitem{Preble2005}{S.~Preble, Q.~Xu, B.~Schmidt, and M.~Lipson, ``Ultrafast all-optical modulation on a silicon chip.'' {\it Opt. Lett.} vol.~30, pp.~2891--2893, 2005.}
\bibitem{Bristow2007}{A.~D.~Bristow, N.~Rotenberg, and H.~M.~Van~Driel, ``Two-photon absorption and Kerr coefficients of silicon for 850-2200 nm.'' {\it Appl.\ Phys.\ Lett.} vol.~90, no.~191104, 2007.}
\bibitem{Wang2013}{T.~Wang, N.~Venkatram, J.~Gosciniak, Y.~Cui et al., ``Multi-photon absorption and third-order nonlinearity in silicon at mid-infrared wavelengths.'' {\it Opt.\ Express} vol.~21, no.~26, pp.~32192--32198, 2013.}
\bibitem{Foster2007}{M.~A.~Foster and A.~L.~Gaeta, ``Wavelength dependence of the ultrafast third-order nonlinearity of silicon,'' presented at Conference on Lasers and Electro-Optics, San Jose, CA, USA, May 6--11, 2007, Paper CTuY5.}
\bibitem{Gai2013}{X.~Gai, Y.~Yu, B.~Kuyken, P.~Ma et al., ``Nonlinear absorption and refraction in crystalline silicon in the mid-infrared.'' {\it Laser Phot.\ Rev.} vol.~7, no.~6, pp.~1054--1064, 2013.}
\bibitem{Lin2007b}{Q.~Lin. J.~Zhang, G.~Piredda, R.~W.~Boyd, P.~M.~Fauchet, and G.~P.~Agrawal, ``Dispersion of silicon nonlinearities in the near infrared region.'' {\it Appl.\ Phys.\ Lett.} vol.~91, no.~021111, 2007.}
\bibitem{Garcia2006}{H.~Garcia and R.~Kalyanaraman, ``Phonon-assisted two-photon absorption in the presence of a dc-field: the nonlinear Franz-Keldysh effect in indirect gap semiconductors.'' {\it J.\ Phys.\ B-At.\ Mol.\ Opt.} vol.~39, no.~12, pp.~2937--2746, 2006.}
\bibitem{Hon2011}{N.~K.~Hon, R.~Soref, and B.~Jalali, ``The third-order nonlinear optical coefficients of Si, Ge, and Si$_{1-x}$Ge$_{x}$ in the midwave and longwave infrared.'' {\it J.\ Appl.\ Phys.} vol.~110, no.~011301, 2011.}
\end{thebibliography}
\end{document}